\documentclass[twocolumn]{jpsj3}

\usepackage{color}
\usepackage{bm}

\title{%
Regularized Continuum Model of a Weyl Semimetal
for Describing Anomalous Electromagnetic Response
}

\author{%
Yositake Takane
}

\inst{%
Department of Quantum Matter, Graduate School of Advanced Sciences of Matter,\\
Hiroshima University, Higashihiroshima, Hiroshima 739-8530, Japan
}

\recdate{ \hspace{50mm} }

\abst{%
Although the Weyl model with an unbounded linear energy spectrum appropriately
describes low-energy electron states in a Weyl semimetal, it cannot capture
the anomalous electromagnetic response of the chiral magnetic effect (CME)
and anomalous Hall effect (AHE) in a straightforward manner.
Here, we propose a regularized continuum model by modifying the Weyl model
and show that it properly describes the CME and AHE in a unified manner.
It turns out that the absence of the CME at equilibrium is
guaranteed by a basic nature of the Berry curvature.
We also show that the original Weyl model can properly describe the CME
if an energy cutoff procedure is appropriately applied,
although it fails to describe the AHE in its present form.
}

\begin{document}
\sloppy
\maketitle

\section{Introduction}

A Weyl semimetal is a three-dimensional gapless system possessing a pair of,
or pairs of, nondegenerate Dirac cones with opposite chirality.~\cite{shindou,
murakami,wan,yang,burkov1,burkov2,imura,WK,delplace,halasz,sekine}
Each Dirac cone is basically described by the Weyl model
with an unbounded linear energy spectrum,
and its band touching point is called the Weyl node.
The $+$ ($-$) chirality of Weyl nodes corresponds to
the monopole (antimonopole) of the Berry curvature with unit strength.
This topological character gives rise to the unusual electromagnetic response
of the anomalous Hall effect (AHE)~\cite{burkov1}
and chiral magnetic effect (CME).~\cite{nielsen,zyuzin}
The AHE designates the phenomenon that a Hall effect is induced
in the absence of an external magnetic field.
The CME represents the anomalous induction of a charge current
in response to an external magnetic field.~\cite{nielsen,zyuzin,fukushima,
son1,son2,basar,liu,vazifeh,chen,goswami,landsteiner,chang,takane1,baireuther}
Our attention is focused on the CME
in the presence of only an external magnetic field.
To date, some materials have been experimentally identified
as Weyl semimetals.~\cite{weng,huang1,xu1,lv1,lv2,xu2,souma,kuroda}

We hereafter focus on a Weyl semimetal with a pair of Weyl nodes
with opposite chiralities assuming that the $+$ and $-$ nodes are respectively
located at $(\mib{k}, E) = (\mib{k}_{R},b_{0})$ and $(\mib{k}_{L}, -b_{0})$
in reciprocal and energy space,
where $\mib{k}_{R} = (0,0,k_{0})$ and $\mib{k}_{L} = (0,0,-k_{0})$.
The AHE is simply caused by the chiral surface states,~\cite{wan}
which appear on a surface to connect the Weyl nodes
in the corresponding surface Brillouin zone.
They cannot appear on the $xy$ plane as the Weyl nodes are projected
onto the identical point of $(k_{x},k_{y}) = (0,0)$
in the surface Brillouin zone.
That is, they typically appear on the $xz$ and $yz$ planes.
In the presence of an external electric field in the $x$- or $y$-direction,
the anomalous Hall response appears in the direction perpendicular to
the external field as long as $k_{0} \neq 0$.
Contrastingly, the behavior of the CME has been a point of controversy.
An early study~\cite{zyuzin} based on the Weyl model predicted that
a finite charge current due to the CME appears even at equilibrium
in response to a static magnetic field if $b_{0} \neq 0$.
Several authors~\cite{vazifeh,chen,chang} have examined this result
by using a lattice model and concluded that the CME current vanishes
under a static magnetic field at equilibrium but can appear
in nonequilibrium situations with a time-dependent magnetic field.

The appearance of the CME at equilibrium should be regarded as an artifact
arising from pathological features of the Weyl model
or their inappropriate regularization.
Related to this, it is worth noting that if we analyze the CME
by using the Weyl model with a linear response theory,~\cite{takane1}
we erroneously find that a finite charge current appears even at equilibrium
[see Eq.~(\ref{eq:error})].
A difficulty also arises if the Weyl model is applied to analyze the AHE.
Indeed, the Weyl model itself cannot describe the chiral surface states,
which are the very origin of the AHE.
As the Weyl model appropriately describes low-energy electron states near
the Weyl nodes, it is natural to consider that its oversimplified
energy spectrum is responsible for the difficulties mentioned above.
However, this consideration has not been examined in a direct manner.

In this paper, we propose a regularized continuum model
for the Weyl semimetal by modifying the Weyl model.
Although the spectrum of this continuum model is also unbounded,
it properly describes the AHE and CME in a unified manner
without relying on a regularization procedure.
From the analysis based on the concept of the Berry curvature,
we show that the absence of the CME at equilibrium is
guaranteed by a basic nature of the Berry curvature.
We also show that the original Weyl model can properly describe the CME
if an energy cutoff is applied in a careful manner,
although it fails to describe the AHE in its present form.
In the next section, we present the regularized continuum model
with an unbounded spectrum and derive expressions for the Hall conductivity,
$\sigma^{\rm AHE}$, and the coefficient for the CME, $\alpha^{\rm CME}$.
In Sect.~3, we analytically calculate $\sigma^{\rm AHE}$ and $\alpha^{\rm CME}$
with the regularized continuum model and show that the results are
consistent with those of a lattice model.
That is, the regularized continuum model properly describes the AHE and CME.
The difficulty of the original Weyl model in capturing the AHE
is clarified in the process of this calculation.
In Sect.~4, we show that the CME can be properly described
by the Weyl model under an energy cutoff.
The last section is devoted to a summary.
We set $\hbar = k_{\rm B} = 1$ throughout this paper.

\section{Model and Formulation}

Hereafter, the energy valley associated with the $+$ node at
$(\mib{k},E) = (\mib{k}_{R},b_{0})$ is referred to as
the right valley, while that associated with the $-$ node at
$(\mib{k}_{L},-b_{0})$ is referred to as the left valley.
We use $\zeta$ to specify the right and left valleys:
$\zeta = R$ for the right valley and $\zeta = L$ for the left valley.
Let us introduce the Weyl model, which is composed of the Weyl Hamiltonians
$H_{R}^{0}$ and $H_{L}^{0}$ for the right and left valleys:
\begin{align}
        \label{eq:Weyl-H0+}
   H_{R}^{0}
 & = v\left[\sigma_{x}k_{x}+\sigma_{y}k_{y}+\sigma_{z}(k_{z}-k_{0})\right]
       + b_{0} -\mu ,
        \\
        \label{eq:Weyl-H0-}
   H_{L}^{0}
 & = v\left[-\sigma_{x}k_{x}-\sigma_{y}k_{y}+\sigma_{z}(-k_{z}-k_{0})\right]
       - b_{0} - \mu ,
\end{align}
where $v$ and $\mu$ respectively denote the velocity and chemical potential,
and $\sigma_{a}$ with $a = x,y,z$ are the Pauli matrices.

Let us introduce a regularized continuum model by modifying the Weyl model.
We respectively replace $\sigma_{z}v(k_{z}-k_{0})$ in $H_{R}^{0}$ and
$\sigma_{z}v(-k_{z}-k_{0})$ in $H_{L}^{0}$
with $\sigma_{z}\Delta_{R}(k_{z})$ and $\sigma_{z}\Delta_{L}(k_{z})$,
and $b_{0}$ in $H_{R}^{0}$ and $-b_{0}$ in $H_{L}^{0}$
with $\Gamma_{R}(k_{z})$ and $\Gamma_{L}(k_{z})$ (see Fig.~1).
In addition, we include $\sigma_{z}\Lambda(k_{x},k_{y})$
in $H_{R}^{0}$ and $-\sigma_{z}\Lambda(k_{x},k_{y})$ in $H_{L}^{0}$ with
\begin{align}
   \Lambda(k_{x},k_{y}) = B(k_{x}^{2}+k_{y}^{2}) ,
\end{align}
where $B$ is positive and very small.
The mass terms with $\sigma_{z}\Lambda$ not only ensure the convergence of
integration over $k_{x}$ and $k_{y}$
but also determine the appearance of chiral surface states.
For the convergence of integration over $k_{z}$,
we restrict $k_{z}$ within the interval of $[-k_{M},k_{M}]$.
Consequently, the Hamiltonians are rewritten as
\begin{align}
        \label{eq:Weyl-H+}
   H_{R}
 & = v\left(\sigma_{x}k_{x}+\sigma_{y}k_{y}\right)
            +\sigma_{z}\left[\Delta_{R}(k_{z})+\Lambda(k_{x},k_{y})\right]
     \nonumber \\
 &   \hspace{10mm}
     + \Gamma_{R}(k_{z})-\mu ,
        \\
        \label{eq:Weyl-H-}
   H_{L}
 & = -v\left(\sigma_{x}k_{x}+\sigma_{y}k_{y}\right)
            +\sigma_{z}\left[\Delta_{L}(k_{z})-\Lambda(k_{x},k_{y})\right]
     \nonumber \\
 &   \hspace{10mm}
     + \Gamma_{L}(k_{z})-\mu ,
\end{align}
which compose the regularized continuum model.
We assume that $\Delta_{R}(k_{z})$ [$\Delta_{L}(k_{z})$] is a monotonically
increasing (decreasing) function of $k_{z}$ satisfying
$\Delta_{R}(k_{z}) \approx v(k_{z}-k_{0})$
for $k_{z} \in (k_{0}-k_{W},k_{0}+k_{W})$,
$\Delta_{L}(k_{z}) \approx v(-k_{z}-k_{0})$
for $k_{z} \in (-k_{0}-k_{W},-k_{0}+k_{W})$, and
\begin{align}
     \label{eq:cond-1}
  \Delta_{R}(k_{M})
  & = \Delta_{L}(-k_{M}) \equiv \Delta_{p} ,
      \\
     \label{eq:cond-2}
  \Delta_{L}(k_{M})
  & = \Delta_{R}(-k_{M}) \equiv -\Delta_{n}
\end{align}
with $\Delta_{p}$, $\Delta_{n} > 0$.
We also assume that $\Gamma_{\zeta}(k_{z})$ satisfies
$\Gamma_{R}(k_{z}) \approx b_{0}$
for $k_{z} \in (k_{0}-k_{W},k_{0}+k_{W})$,
$\Gamma_{L}(k_{z}) \approx -b_{0}$
for $k_{z} \in (-k_{0}-k_{W},-k_{0}+k_{W})$, and
\begin{align}
     \label{eq:cond-3}
  \Gamma_{R}(\pm k_{M}) =
  \Gamma_{L}(\pm k_{M}) = 0 .
\end{align}
The simplest choice of $\Delta_{\zeta}$ is
$\Delta_{R} = v(k_{z}-k_{0})$ and $\Delta_{L} = v(-k_{z}-k_{0})$,
the same as those in the original Weyl model,
leading to $\Delta_{p} = v(k_{M}-k_{0})$ and $\Delta_{n} = v(k_{M}+k_{0})$.
We observe in Sect.~3 that the anomalous electromagnetic response
does not depend on explicit functional forms of
$\Delta_{\zeta}(k_{z})$ and $\Gamma_{\zeta}(k_{z})$, indicating that
it is governed by the topological character of a Weyl semimetal.
\begin{figure}[btp]
\begin{center}
\includegraphics[height=2.8cm]{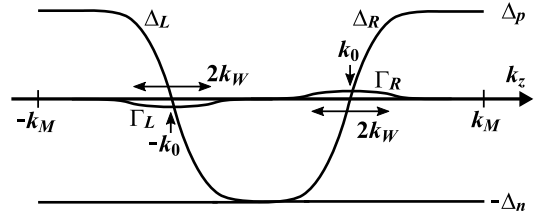}
\end{center}
\caption{
$k_{z}$ dependences of $\Delta_{R}(k_{z})$, $\Delta_{L}(k_{z})$,
$\Gamma_{R}(k_{z})$, and $\Gamma_{L}(k_{z})$.
}
\end{figure}

A comment on the mass terms with $\sigma_{z}\Lambda$ is in order.
They ensure the appearance of chiral surface states~\cite{takane2,takane3}
as well as the convergence of integration over $k_{x}$ and $k_{y}$.
Indeed, if $B$ in $\Lambda$ is set equal to zero,
$H_{\zeta}$ can support no chiral surface states.
The sign of $B$ determines in what region of $k_{z}$ the chiral surface states
appear when the system of a Weyl semimetal has
a surface not perpendicular to the $z$-axis.
With these terms in Eqs.~(\ref{eq:Weyl-H+}) and (\ref{eq:Weyl-H-}),
$H_{R}$ supports the chiral surface states with $+$ chirality
in the region of $-k_{M} < k_{z} < k_{0}$,
while $H_{L}$ supports those with $-$ chirality in the region of
$-k_{M} < k_{z} < -k_{0}$.
Hence, the chiral surface states with both $+$ and $-$ chiralities are present
when $-k_{M} < k_{z} < -k_{0}$,
indicating that this region is topologically trivial.
Thus, we observe that only the region of $-k_{0} < k_{z} < k_{0}$
is topologically nontrivial.
The AHE should be induced by the electron states in this region.
Note that the region of $-k_{0} < k_{z} < k_{0}$ remains topologically
nontrivial even if the sign of the mass terms is reversed.
However, after the sign reversal, $H_{R}$ supports the chiral surface states
with $-$ chirality in the region of $k_{0} < k_{z} < k_{M}$,
while $H_{L}$ supports those with $+$ chirality
in the region of $-k_{0} < k_{z} < k_{M}$.

We introduce the Berry curvature, which plays an essential role
in describing the AHE and CME,
and then derive expressions for the Hall conductivity and the coefficient
of the CME in terms of the Berry curvature.
Let us define $\mib{d}_{\zeta}(\mib{k})$ as
\begin{align}
   \mib{d}_{R}(\mib{k})
     & = \bigl(vk_{x}, vk_{y}, \Delta_{R}(k_{z})+\Lambda(k_{x},k_{y})\bigr) ,
   \\
   \mib{d}_{L}(\mib{k})
     & = \bigl(-vk_{x}, -vk_{y}, \Delta_{L}(k_{z})-\Lambda(k_{x},k_{y})\bigr) .
\end{align}
We also define $d_{\zeta}(\mib{k})=|\mib{d}_{\zeta}(\mib{k})|$ and
\begin{align}
      \label{eq:def-d-hat}
   \hat{\mib{d}}_{\zeta}(\mib{k})
      = \frac{\mib{d}_{\zeta}(\mib{k})}{d_{\zeta}(\mib{k})} .
\end{align}
The Berry curvatures, $\Omega^{a}_{\zeta}(\mib{k})$ with $a = x,y,z$,
are defined as
\begin{align}
  \Omega^{x}_{\zeta}(\mib{k})
    & = -\frac{1}{2d_{\zeta}^{3}}\mib{d}_{\zeta}\cdot
        \left(\frac{\partial \mib{d}_{\zeta}}{\partial k_{z}}
              \times
              \frac{\partial \mib{d}_{\zeta}}{\partial k_{y}}\right) ,
           \\
  \Omega^{y}_{\zeta}(\mib{k})
    & = -\frac{1}{2d_{\zeta}^{3}}\mib{d}_{\zeta}\cdot
        \left(\frac{\partial \mib{d_{\zeta}}}{\partial k_{x}}
              \times
              \frac{\partial \mib{d}_{\zeta}}{\partial k_{z}}\right) ,
           \\
  \Omega^{z}_{\zeta}(\mib{k})
    & = -\frac{1}{2d_{\zeta}^{3}}\mib{d}_{\zeta}\cdot
        \left(\frac{\partial \mib{d}_{\zeta}}{\partial k_{y}}
              \times
              \frac{\partial \mib{d}_{\zeta}}{\partial k_{x}}\right) .
\end{align}
In each valley, the energy of the state with $\mib{k}$ measured from $\mu$
is expressed as
\begin{align}
  E^{\eta}_{\zeta}(\mib{k})
     = \Gamma_{\zeta}(k_{z})-\mu + \eta d_{\zeta}(\mib{k}) ,
\end{align}
where $\eta = +$ for the conduction band and
$\eta = -$ for the valence band (see Fig.~2).
\begin{figure}[btp]
\begin{center}
\includegraphics[height=3.4cm]{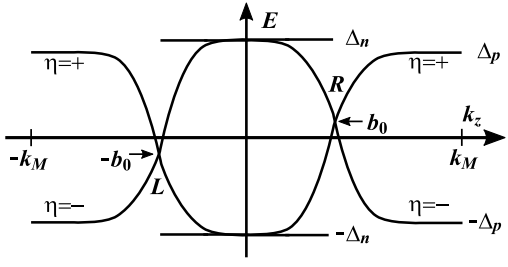}
\end{center}
\caption{
Energy dispersion as a function of $k_{z}$ for $(k_{x},k_{y})=(0,0)$, where
$\eta = +$ and $-$ respectively represent the conduction and valence bands.
}
\end{figure}

The charge current $j^{a}$ is decomposed as
\begin{align}
       \label{eq:def-j-a}
   j^{a} = j_{R}^{a} + j_{L}^{a} ,
\end{align}
where $j_{R}^{a}$ and $j_{L}^{a}$ respectively represent the contributions from
the right and left valleys:
\begin{align}
       \label{eq:def-current}
  j_{\zeta}^{a}
  = - e \frac{\partial\left(\Gamma_{\zeta}
                            +\mib{\sigma}\cdot\mib{d}_{\zeta}\right)}
             {\partial k_{a}} .
\end{align}
Within a linear response theory, the average current induced
by a vector potential $\mib{A} = (A_{x},A_{y},A_{z})$ is expressed as
\begin{align}
  \langle j_{\zeta}^{a} \rangle
 & = -\Pi_{\zeta}^{ab}(\mib{q},\omega)A_{b}(\mib{q},\omega) .
\end{align}
The response function $\Pi_{\zeta}^{ab}$ can be obtained by using
the analytic continuation of $i\nu \to \omega+i\delta$
from its Matsubara representation,
\begin{align}
           \label{eq:def-response}
    \Pi_{\zeta}^{ab}(\mib{q},i\nu)
  &  = e^{2} \int\frac{d^{3}k}{(2\pi)^{3}}\frac{1}{\beta}\sum_{\epsilon}
       {\rm Tr} \Biggl\{\frac{\partial\left(\Gamma_{\zeta}
                          +\mib{\sigma}\cdot\mib{d}_{\zeta}\right)}
                               {\partial k_{a}}
          \nonumber \\
  &    \hspace{-14mm} \times
             G_{\zeta}(\mib{k}+\mib{q},i\epsilon+i\nu)
             \frac{\partial\left(\Gamma_{\zeta}
                                 +\mib{\sigma}\cdot\mib{d}_{\zeta}\right)}
                  {\partial k_{b}}
             G_{\zeta}(\mib{k},i\epsilon)\Biggr\} ,
\end{align}
where $\beta$ is the inverse of temperature $T$.
Here, the thermal Green's function is given by
\begin{align}
       \label{eq:def-GF}
  G_{\zeta}(\mib{k},i\epsilon)
  = \sum_{\eta = \pm}
    \frac{1}{2}\bigl(1+\eta\mib{\sigma}\cdot\hat{\mib{d}}_{\zeta}\bigr)
    \frac{1}{i\epsilon-E_{\zeta}^{\eta}} .
\end{align}

The following derivation of the transport coefficients basically
relies on the approach of Chang and Yang.~\cite{chang}
Let us consider the AHE.
We derive an expression for $\Pi_{\zeta}^{yx}$, which
determines the Hall current $\langle j^{y}\rangle$ in the $y$-direction
induced by an electric field $\mib{E}=(E_{x},0,0)$, where we assume
that $\mib{E}(\mib{q},\omega) = i\omega \mib{A}(\mib{q},\omega)$
with $\mib{A}=(A_{x},0,0)$.
The Hall current is expressed as
$\langle j^{y} \rangle = \sigma^{\rm AHE} E_{x}$
with $\sigma^{\rm AHE}$ being the Hall conductivity.
In the uniform limit of $\mib{q} = \mib{0}$,
the Hall conductivity is given by
\begin{align}
    \sigma^{\rm AHE}(\omega)
    = - \frac{1}{i\omega}\Pi^{yx}(\mib{0},i\nu)\big|_{i\nu \to \omega+i\delta}
\end{align}
with $\Pi^{yx}\equiv\Pi_{R}^{yx}+\Pi_{L}^{yx}$.
Performing the trace and the Matsubara summation, we find
\begin{align}
      \label{eq:Pi_yx}
  \Pi_{\zeta}^{yx}(\mib{0},i\nu)
  & = e^{2}\int\frac{d^{3}k}{(2\pi)^{3}} \,
      \Omega_{\zeta}^{z}
           \nonumber \\
  & \hspace{-7mm} \times
      \frac{-4\nu d_{\zeta}^{2}}{4d_{\zeta}^{2}+\nu^{2}}
      \left(f_{\rm FD}(E_{\zeta}^{+})-f_{\rm FD}(E_{\zeta}^{-}) \right),
\end{align}
where $f_{\rm FD}$ denotes the Fermi--Dirac function
\begin{align}
  f_{\rm FD}(E) = \frac{1}{e^{\beta E}+1} .
\end{align}
Concerning the AHE, we hereafter focus on the simplest case
where $\Gamma_{\zeta} = 0$ (i.e., $b_{0} = 0$)
and $\mu$ is located at the Weyl nodes (i.e., $\mu = 0$).
Then, we find from Eq.~(\ref{eq:Pi_yx}) that $\sigma^{\rm AHE}$
in the static limit of $\omega \to 0$ is expressed as
\begin{align}
        \label{eq:sigma^AHE}
    \sigma^{\rm AHE}_{\rm stat}
    = e^{2} \int\frac{d^{3}k}{(2\pi)^{3}}\sum_{\zeta= R, L}
      \Omega_{\zeta}^{z}(\mib{k}) .
\end{align}

Let us turn to the CME.
We derive an expression for $\Pi_{\zeta}^{zy}$,
which determines the CME current $\langle j^{z}\rangle$ in the $z$-direction
induced by a magnetic field $\mib{B}=(0,0,B_{z})$,
where $\mib{B}(\mib{q},\omega) = i\mib{q}\times\mib{A}(\mib{q},\omega)$
with $\mib{A}=(0,A_{y},0)$ and $\mib{q} = (q,0,0)$.
The CME current is expressed as
$\langle j^{z} \rangle = \alpha^{\rm CME}B_{z}$, where
\begin{align}
      \label{eq:CME-C}
   \alpha^{\rm CME}(\mib{q},\omega)
   = - \frac{1}{iq}\Pi^{zy}(\mib{q},i\nu)\big|_{i\nu \to \omega+i\delta}
\end{align}
with $\Pi^{zy}\equiv\Pi_{R}^{zy}+\Pi_{L}^{zy}$.
Substituting Eq.~(\ref{eq:def-GF}) into Eq.~(\ref{eq:def-response}),
we observe that each term of the resulting expression for $\Pi_{\zeta}^{zy}$
contains $\eta = \pm $ in $G_{\zeta}(\mib{k},i\epsilon)$ and
$\eta' = \pm$ in $G_{\zeta}(\mib{k}+\mib{q},i\epsilon+i\nu)$.
We separately treat the intraband contribution $\Pi_{\rm intra}^{zy}$
arising from the terms with $\eta = \eta'$ and the interband contribution
$\Pi_{\rm inter}^{zy}$ arising from the terms with $\eta \neq \eta'$.
After a lengthy calculation, the intraband contribution
for a small $q$ is expressed as
\begin{align}
  \Pi_{\rm intra}^{zy}(\mib{q},i\nu)
  & = e^{2}\int\frac{d^{3}k}{(2\pi)^{3}}\sum_{\zeta = R,L}\sum_{\eta=\pm}
      (iq)d_{\zeta}
          \nonumber \\
  & \hspace{-16mm}\times
      \left( \Omega_{\zeta}^{y}
             \frac{\partial E_{\zeta}^{\eta}}{\partial k_{y}}
           + \Omega_{\zeta}^{z}
             \frac{\partial E_{\zeta}^{\eta}}{\partial k_{z}}
      \right)
      \frac{\partial f_{\rm FD}(E_{\zeta}^{\eta})}{\partial E_{\zeta}^{\eta}}
      \frac{-q\frac{\partial E_{\zeta}^{\eta}}{\partial k_{x}}}
           {i\nu-q\frac{\partial E_{\zeta}^{\eta}}{\partial k_{x}}} .
\end{align}
The interband contribution for a small $q$ is expressed as
\begin{align}
  \Pi_{\rm inter}^{zy}(\mib{q},i\nu)
  & = e^{2}\int\frac{d^{3}k}{(2\pi)^{3}}\sum_{\zeta = R,L}\sum_{\eta=\pm}
      (iq)
          \nonumber \\
  & \hspace{-14mm}\times
      \Biggl[
       - d_{\zeta}
         \left\{
            \eta
            \left( \Omega_{\zeta}^{y}
                   \frac{\partial d_{\zeta}}{\partial k_{y}}
                 + \Omega_{\zeta}^{z}
                   \frac{\partial d_{\zeta}}{\partial k_{z}}
            \right)
          + \Omega_{\zeta}^{z}\frac{\partial \Gamma_{\zeta}}{\partial k_{z}}
         \right\}
         \Upsilon
            \nonumber \\
  & \hspace{5mm}
        - 2 \eta  d_{\zeta}^{2} \Omega_{\zeta}^{x} 
          \frac{\partial E_{\zeta}^{-\eta}}{\partial k_{x}}
          \frac{\partial \Upsilon}
               {\partial E_{\zeta}^{-\eta}}
      \Biggr] ,
\end{align}
where
\begin{align}
   \Upsilon
   = \frac{f_{\rm FD}(E_{\zeta}^{\eta})-f_{\rm FD}(E_{\zeta}^{-\eta})}
          {i\nu + E_{\zeta}^{\eta} - E_{\zeta}^{-\eta}} .
\end{align}

In the static limit of $\omega \to 0$ before $q \to 0$,
the coefficient of the CME is given by adding the intraband and interband
contributions as
\begin{align}
        \label{eq:CME-stat-pre}
  \alpha^{\rm CME}_{\rm stat}
  & = e^{2}\int\frac{d^{3}k}{(2\pi)^{3}}\sum_{\zeta = R,L}\sum_{\eta=\pm}
        \nonumber \\
  & \hspace{-14mm}
      \times
      \Biggl[- d_{\zeta}\mib{\Omega}_{\zeta}\cdot\nabla_{\mib k}
               E_{\zeta}^{\eta}
               \frac{\partial f_{\rm FD}(E_{\zeta}^{\eta})}
                    {\partial E_{\zeta}^{\eta}}
             + \eta\Omega_{\zeta}^{z}
               \frac{\partial \Gamma_{\zeta}}{\partial k_{z}}
               f_{\rm FD}(E_{\zeta}^{\eta})
      \Biggr] .
\end{align}
Performing the partial integration over $\mib{k}$ using
$\nabla_{\mib k}\cdot\mib{\Omega}_{R} = 2\pi\delta(\mib{k}-\mib{k}_{R})$
and $\nabla_{\mib k}\cdot\mib{\Omega}_{L} = - 2\pi\delta(\mib{k}-\mib{k}_{L})$,
we arrive at
\begin{align}
        \label{eq:CME-stat}
  \alpha^{\rm CME}_{\rm stat}
    = e^{2}\sum_{\zeta = R,L}\sum_{\eta=\pm}
      \int\frac{d^{3}k}{(2\pi)^{3}}
      \eta \mib{\Omega}_{\zeta} \cdot
      \left(\nabla_{\mib{k}}E_{\zeta}^{\eta}\right)
      f_{\rm FD}(E_{\zeta}^{\eta}) .
\end{align}
In deriving Eq.~(\ref{eq:CME-stat}), we ignore the surface term given by
\begin{align}
  c_{S}
   = \lim_{k_{r} \to \infty}\frac{-e^{2}}{(2\pi)^{3}}
     \int_{S}dS_{\mib{k}}\sum_{\zeta=R,L}\sum_{\eta=\pm} \mib{n}\cdot
     d_{\zeta}\mib{\Omega}_{\zeta}f_{\rm FD}(E_{\zeta}^{\eta}) ,
\end{align}
where $S$ represents the cylindrical surface
of height $2k_{M}$ and radius $k_{r}$,
and $\mib{n}$ is the outward unit vector normal to this surface.
In the regularized continuum model, we can show that $c_{S}= 0$
for an arbitrary $\mu$ under the conditions of
$\Delta_{R}(\pm k_{M}) = \Delta_{L}(\mp k_{M})$ and
$\Gamma_{R}(\pm k_{M}) = \Gamma_{L}(\mp k_{M})$,
where $\pm \sigma_{z}\Lambda$ plays the role of a convergence factor.
However, $c_{S}$ does not vanish in the original Weyl model; hence,
caution is necessary in applying Eq.~(\ref{eq:CME-stat}) to the Weyl model.
We argue this in Sect.~4.

Let us finally consider the weakly nonequilibrium situation with
an external magnetic field that is spatially uniform and
slowly oscillating in the time domain.
The response for it is characterized by $\alpha^{\rm CME}_{\rm neq}$
defined in the limit of $q \to 0$ before $\omega \to 0$.
The intraband contribution vanishes in this limit
and then $\alpha^{\rm CME}_{\rm neq}$ is expressed as
\begin{align}
        \label{eq:CME-neq}
  \alpha^{\rm CME}_{\rm neq}
  & = e^{2}\int\frac{d^{3}k}{(2\pi)^{3}}\sum_{\zeta = R,L}\sum_{\eta=\pm}
          \nonumber \\
  & \hspace{-6mm}\times
      \Biggl[
        - d_{\zeta}\Omega_{\zeta}^{x}
          \frac{\partial E_{\zeta}^{\eta}}{\partial k_{x}}
          \frac{\partial f_{\rm FD}(E_{\zeta}^{\eta})}
               {\partial E_{\zeta}^{\eta}}
        + \eta\Omega_{\zeta}^{z}\frac{\partial \Gamma_{\zeta}}{\partial k_{z}}
          f_{\rm FD}(E_{\zeta}^{\eta})
      \Biggr].
\end{align}

\section{Anomalous Electromagnetic Response}

In this section, we show that the regularized continuum model
properly describes the AHE and CME in a unified manner.
For simplicity, we restrict our consideration to zero temperature,
at which $f_{\rm FD}(E) = \theta(-E)$
with $\theta(x)$ being the Heaviside step function.

\subsection{Anomalous Hall effect}

We start with the expression for the Hall conductivity,
Eq.~(\ref{eq:sigma^AHE}), in the static limit
under the assumption of $\Gamma_{\zeta} = 0$ and  $\mu = 0$.
Let us consider $\Phi_{\zeta}(k_{z})$ defined by
\begin{align}
    \Phi_{\zeta}(k_{z})
    = \frac{1}{2\pi}
      \int_{-\infty}^{\infty}dk_{x}\int_{-\infty}^{\infty}dk_{y} \,
      \Omega_{\zeta}^{z}(\mib{k}) ,
\end{align}
in terms of which $\sigma^{\rm AHE}_{\rm stat}$ is expressed as
\begin{align}
    \sigma^{\rm AHE}_{\rm stat}
    = \frac{e^{2}}{(2\pi)^{2}}\int_{-k_{M}}^{k_{M}}dk_{z}
      \sum_{\zeta=R,L} \Phi_{\zeta}(k_{z}) .
\end{align}
From the definition of $\Omega_{\zeta}^{z}$, we observe that
$\Phi_{\zeta}(k_{z})$ represents a winding number that counts how many times
$\hat{\mib{d}}_{\zeta}$, defined in Eq.~(\ref{eq:def-d-hat}), wraps
the unit sphere when $(k_{x},k_{y})$ moves all over the two-dimensional space.
Note that $\sum_{\zeta=R,L}\Phi_{\zeta}(k_{z})$ plays the role of
the Chern number at a given $k_{z}$.
The $z$ component of the Berry curvature is expressed as
\begin{align}
  \Omega_{R}^{z}(\mib{k})
  & = \frac{v^{2}}{2d_{R}^{3}}
      \left[\Delta_{R}(k_{z})-\Lambda(k_{x},k_{y})\right] ,
      \\
  \Omega_{L}^{z}(\mib{k})
  & = \frac{v^{2}}{2d_{L}^{3}}
      \left[\Delta_{L}(k_{z})+\Lambda(k_{x},k_{y})\right] .
\end{align}
Performing the integration over $k_{x}$ and $k_{y}$ in terms of
$k_{\perp}=\sqrt{k_{x}^{2}+k_{y}^{2}}$, we easily find that
\begin{align}
    \label{eq:Phi-R}
  \Phi_{R}(k_{z})
  & = \left.\frac{-\left[\Delta_{R}(k_{z})+\Lambda(k_{\perp})\right]}
                 {2 \sqrt{v^{2}k_{\perp}^{2}
                      + \left[\Delta_{R}(k_{z})+\Lambda(k_{\perp})\right]^{2}}}
      \right|_{k_{\perp}=0}^{k_{\perp}=\infty}
    \nonumber \\
  & = \frac{1}{2}\bigl(-1+{\rm sign}\{\Delta_{R}(k_{z})\}\bigr) ,
    \\
    \label{eq:Phi-L}
  \Phi_{L}(k_{z})
  & = \left.\frac{-\left[\Delta_{L}(k_{z})-\Lambda(k_{\perp})\right]}
                 {2 \sqrt{v^{2}k_{\perp}^{2}
                      + \left[\Delta_{L}(k_{z})-\Lambda(k_{\perp})\right]^{2}}}
      \right|_{k_{\perp}=0}^{k_{\perp}=\infty} .
    \nonumber \\
  & = \frac{1}{2}\bigl(1+{\rm sign}\{\Delta_{L}(k_{z})\}\bigr) .
\end{align}
Note that $\Phi_{\zeta}(k_{z})$ takes an integer
owing to the presence of $\Lambda$.
We observe that $\sum_{\zeta=R,L}\Phi_{\zeta}(k_{z}) = -1$ in the region of
$-k_{0} < k_{z} < k_{0}$ and vanishes otherwise.
We finally find that
\begin{align}
       \label{eq:result-AHE}
    \sigma^{\rm AHE}_{\rm stat}
    = - \frac{e^{2}}{2\pi^{2}}k_{0} ,
\end{align}
which is equivalent to the result reported previously.~\cite{burkov1}

Although Eq.~(\ref{eq:result-AHE}) is not changed
if $\Lambda$ is set equal to zero,~\cite{ludwig}
the presence of $\Lambda$ is essential in describing the AHE
since the model with $\Lambda = 0$ cannot support the chiral surface states.
Generally speaking, a continuum model can describe the AHE
including the chiral surface states
only when $\Phi_{\zeta}(k_{z})$ takes an integer value in each valley and
$\sum_{\zeta=R,L}\Phi_{\zeta}(k_{z})$ is identified with the Chern number.
If the sign of the mass terms with $\sigma_{z}\Lambda$
is reversed in $H_{\zeta}$, the final result is not affected,
although $\Phi_{\zeta}(k_{z})$ is changed.
This is consistent with the argument in Sect.~2.

\subsection{Chiral magnetic effect in the static limit}

We turn to the CME at equilibrium, which is characterized by
the coefficient of the CME in the static limit, $\alpha^{\rm CME}_{\rm stat}$,
given by Eq.~(\ref{eq:CME-stat}).
Since the CME does not appear at equilibrium,~\cite{vazifeh,chang}
$\alpha^{\rm CME}_{\rm stat}$ must vanish
for the proper description of the CME.
We show that $\alpha^{\rm CME}_{\rm stat} = 0$ within the framework of
the regularized continuum model.
To do so, it is insightful to consider the contribution to
$\alpha^{\rm CME}_{\rm stat}$ arising from
the energy interval of $(E,E+\Delta E)$.
This is expressed as
\begin{align}
  \Delta\alpha^{\rm CME}_{\rm stat}
  = \frac{\partial \alpha^{\rm CME}_{\rm stat}}{\partial E}\Delta E ,
\end{align}
where
\begin{align}
  \frac{\partial \alpha^{\rm CME}_{\rm stat}}{\partial E}
  & = e^{2}\int\frac{d^{3}k}{(2\pi)^{3}}\sum_{\zeta = R,L}\sum_{\eta=\pm}
        \nonumber \\
  & \hspace{5mm} \times
    \eta \mib{\Omega}_{\zeta} \cdot \nabla_{\mib k}E_{\zeta}^{\eta}
    \left.\frac{\partial f_{\rm FD}(E_{\zeta}^{\eta})}
               {\partial \mu}\right|_{\mu = E} .
\end{align}
Using $\partial f_{\rm FD}/\partial \mu =
- \partial f_{\rm FD}/\partial E_{\zeta}^{\eta}$ and Gauss's theorem,
we find that
\begin{align}
        \label{eq:par-alpha}
  \frac{\partial \alpha^{\rm CME}_{\rm stat}}{\partial E}
  = -\frac{e^{2}}{(2\pi)^{3}} \Xi(E)
\end{align}
with
\begin{align}
        \label{eq:def-Xi}
  \Xi
  = \lim_{k_{r} \to \infty} \int_{S} dS \,
    \sum_{\zeta = R,L}\sum_{\eta = \pm}
    \left. \eta{\mib n} \cdot \mib{\Omega}_{\zeta}
    f_{\rm FD}(E_{\zeta}^{\eta})\right|_{\mu = E} ,
\end{align}
where $S$ represents the cylindrical surface
of height $2k_{M}$ and radius $k_{r}$.
Note that $\Xi$ inevitably vanishes in a lattice model owing to
the periodicity in the first Brillouin zone,~\cite{comment1}
indicating that the absence of the CME at equilibrium is guaranteed by
the conservation of the Berry curvature in reciprocal space.
In our continuum model, the integration over the side surface of
area $2\pi k_{r}\times 2k_{M}$ vanishes in the limit of $k_{r} \to \infty$;
thus, $\Xi = 0$ is justified if the contributions from
the top and bottom surfaces at $k_{z} = \pm k_{M}$ are cancelled out.
This is ensured when the two valleys are connected with each other through
the virtual boundaries at $k_{z} = \pm k_{M}$ under the conditions of
$\Delta_{R}(\pm k_{M}) = \Delta_{L}(\mp k_{M})$ and
$\Gamma_{R}(\pm k_{M})=\Gamma_{R}(\mp k_{M})$.
That is, the edge of the right valley at $k_{z} = k_{M}$ ($-k_{M}$)
is connected to that of the left valley at $k_{z} = -k_{M}$ ($k_{M}$).
We can show that $\pm \sigma_{z}\Lambda$ plays no role
in the integration over the top and bottom surfaces.
Hence,
\begin{align}
      \label{eq:vanishing-der-alph}
  \frac{\partial \alpha^{\rm CME}_{\rm stat}}{\partial E} = 0
\end{align}
at an arbitrary $E$ under the assumption of
Eqs.~(\ref{eq:cond-1})--(\ref{eq:cond-3}).

Note that the requirement on $\Delta_{\zeta}(\pm k_{M})$
and $\Gamma_{\zeta}(\pm k_{M})$ is necessary to justify $\Xi(E) = 0$
only in the case of $E$ being located far away from the Weyl nodes
as $E < -\Delta_{M}$ or $\Delta_{M} < E$,
where $\Delta_{M} = {\rm min}\{\Delta_{p},\Delta_{n}\}$,
and hence the conical structure of each valley is broken
in the $k_{z}$-direction (see Fig.~2).
Indeed, in the case of $-\Delta_{M} < E < \Delta_{M}$,
where the conical structure is preserved,
we can show that $\Xi= 0$ without the requirement.
In this case, we can reduce Eq.~(\ref{eq:def-Xi}) to~\cite{comment2}
\begin{align}
  \Xi
  = \lim_{k_{r} \to \infty} \int_{S} dS \,
    \sum_{\zeta = R,L} \left(-{\mib n}\right) \cdot \mib{\Omega}_{\zeta}
\end{align}
by setting $f_{\rm FD}(E_{\zeta}^{+}) = 0$ and $f_{\rm FD}(E_{\zeta}^{-}) = 1$,
and immediately conclude $\Xi= 0$ by noting that
$\nabla_{\mib k}\cdot\mib{\Omega}_{R} = 2\pi\delta(\mib{k}-\mib{k}_{R})$
and $\nabla_{\mib k}\cdot\mib{\Omega}_{L} = - 2\pi\delta(\mib{k}-\mib{k}_{L})$.
That is, $\Xi= 0$ is guaranteed by the topological character of
the Berry curvature.

Equation~(\ref{eq:vanishing-der-alph}) strongly indicates
the vanishing of $\alpha^{\rm CME}_{\rm stat}$.
However, since the spectrum of our continuum model is unbounded,
we explicitly show that $\alpha^{\rm CME}_{\rm stat} = 0$
by using Eq.~(\ref{eq:CME-stat}) for safety.
We restrict our argument to the case of $\mu < -\Delta_{M}$.
Once $\alpha^{\rm CME}_{\rm stat} = 0$ is verified for a particular $\mu$,
we can generalize the result to an arbitrary $\mu$ using
Eq.~(\ref{eq:vanishing-der-alph}).
Substituting the explicit expression for $\mib{\Omega}_{\zeta}$
into Eq.~(\ref{eq:CME-stat}), we find that
\begin{align}
      \label{eq:alpha-stat_0}
  \alpha^{\rm CME}_{\rm stat}
  & = \frac{e^{2}}{(2\pi)^{3}}\int_{-k_{M}}^{k_{M}}dk_{z}
      \int_{0}^{\infty}dk_{\perp}2\pi k_{\perp}
          \nonumber \\
  & \hspace{-8mm} \times
       \Biggl[ \left(  \frac{v^{2}}{2d_{R}^{2}}
                       \frac{\partial \Delta_{R}}{\partial k_{z}}
                     - \frac{v^{2}(\Delta_{R}-\Lambda)}{2d_{R}^{3}}
                       \frac{\partial \Gamma_{R}}{\partial k_{z}}
               \right) f_{\rm FD}(E_{R}^{-})
          \nonumber \\
  & \hspace{-6mm}
             + \left(  \frac{v^{2}}{2d_{L}^{2}}
                       \frac{\partial \Delta_{L}}{\partial k_{z}}
                     - \frac{v^{2}(\Delta_{L}+\Lambda)}{2d_{L}^{3}}
                       \frac{\partial \Gamma_{L}}{\partial k_{z}}
               \right) f_{\rm FD}(E_{L}^{-})
       \Biggr] ,
\end{align}
where $f_{\rm FD}(E_{\zeta}^{+}) = 0$ is used.
Performing the integrations over $k_{\perp}$ and then over $k_{z}$
in an explicit manner, we can analytically show that
$\alpha^{\rm CME}_{\rm stat}$ vanishes under the conditions of
$\Delta_{R}(\pm k_{M}) = \Delta_{L}(\mp k_{M})$
and $\Gamma_{R}(\pm k_{M})=\Gamma_{L}(\mp k_{M})$.
Explicit functional forms of $\Delta_{\zeta}(k_{z})$ and
$\Gamma_{\zeta}(k_{z})$ are not necessary to derive this result.

To summarize, $\alpha^{\rm CME}_{\rm stat} = 0$ reflects a conservation law
for the Berry curvature that inevitably holds in a lattice model.
In the region of $-\Delta_{M} < E < \Delta_{M}$, the vanishing of $\Xi(E)$
is guaranteed by the topological character of the Berry curvature,
indicating that the conservation law automatically holds
when the conical structure is well defined in each valley.
The additional condition on $\Delta_{\zeta}(\pm k_{M})$
and $\Gamma_{\zeta}(\pm k_{M})$
is required outside this region far away from the Weyl nodes,
where the conical structure is broken in the $k_{z}$-direction.
In such a region, the Berry curvature should be conserved
between the right and left valleys to ensure $\Xi(E) = 0$
through the virtual boundaries at $k_{z} = \pm k_{M}$.
For the CME, $\pm \sigma_{z}\Lambda$ merely plays the role of
a convergence factor.

Note that the Berry curvature is completely conserved in the Weyl model
as its conical structure is preserved at an arbitrary energy.
This suggests that the Weyl model can properly describe the CME
if the unbounded energy spectrum is regularized in a certain manner.
This possibility is examined in Sect.~4.

\subsection{Chiral magnetic effect at nonequilibrium}

We calculate the coefficient of the CME in the low-frequency limit,
$\alpha^{\rm CME}_{\rm neq}$, by using Eq.~(\ref{eq:CME-neq}).
Noting $\alpha^{\rm CME}_{\rm stat} = 0$ and using
Eq.~(\ref{eq:CME-stat-pre}), we rewrite $\alpha^{\rm CME}_{\rm neq}$ as
\begin{align}
      \label{eq:CME-neq-orig}
  \alpha^{\rm CME}_{\rm neq}
  & = e^{2}\int\frac{d^{3}k}{(2\pi)^{3}}\sum_{\zeta = R,L}\sum_{\eta=\pm}
        \nonumber \\
  & \hspace{4mm}
      \times
      d_{\zeta}
      \left( \Omega_{\zeta}^{y}\frac{\partial E_{\zeta}^{\eta}}{\partial k_{y}}
           + \Omega_{\zeta}^{z}\frac{\partial E_{\zeta}^{\eta}}{\partial k_{z}}
      \right)
      \frac{\partial f_{\rm FD}(E_{\zeta}^{\eta})}
      {\partial E_{\zeta}^{\eta}} .
\end{align}
This indicates that only the electron states at the Fermi level contribute
to $\alpha^{\rm CME}_{\rm neq}$.
When $\mu$ is not far away from the Weyl nodes, this allows us
to calculate $\alpha^{\rm CME}_{\rm neq}$ by replacing the parameters
in the regularized continuum model with those of the original Weyl model.
That is, we can set $\Delta_{R} = v(k_{z}-k_{0})$,
$\Delta_{L} = v(-k_{z}-k_{0})$, $\Gamma_{R} = -\Gamma_{L} = b_{0}$,
and $\Lambda = 0$ in calculating Eq.~(\ref{eq:CME-neq-orig}).
We thus find that
\begin{align}
  \alpha^{\rm CME}_{\rm neq} = \frac{e^{2}}{3\pi^{2}}b_{0} ,
\end{align}
which is equivalent to the result reported in Ref.~\citen{takane1}.
This is also consistent with the result of Refs.~\citen{chang}
and \citen{baireuther} based on a lattice model.

\section{Energy Cutoff in the Weyl Model}

In this section, we examine whether the original Weyl model
can properly describe the CME.
As emphasized in Sect.~3, the vanishing of $\alpha^{\rm CME}_{\rm stat}$
is guaranteed by the conservation law of the Berry curvature
in reciprocal space.
Since the Weyl model conserves the Berry curvature at an arbitrary $E$,
it should give the correct result
if the unbounded linear spectrum is appropriately regularized.

We examine this within an energy cutoff procedure.~\cite{landsteiner}
Let us start with Eq.~(\ref{eq:CME-stat-pre})
since Eq.~(\ref{eq:CME-stat}) is not simply justified for the Weyl model.
If the energy cutoff at $E = - E_{c}$ is straightforwardly applied to
Eq.~(\ref{eq:CME-stat-pre}), it plays no role since the first term represents
the contribution from the Fermi level and the second term disappears
as $\partial \Gamma_{\zeta}/\partial k_{z} = 0$.
Consequently, Eq.~(\ref{eq:CME-stat-pre}) gives the incorrect result
\begin{align}
     \label{eq:error}
  \alpha^{\rm CME}_{\rm stat}
  \overset{?}{=} - \frac{e^{2}}{2\pi^{2}}b_{0} .
\end{align}
This puzzling feature can be resolved by performing
the energy cutoff procedure in a careful manner.

An energy cutoff is naturally introduced into an analysis
in the form of the distribution function of electrons.
In our analysis, the energy cutoff at $E = - E_{c}$ is fully
taken into account by replacing the Fermi--Dirac function
$f_{\rm FD}(E_{\zeta}^{\eta})$ with
\begin{align}
  \tilde{f}_{\rm FD}(E_{\zeta}^{\eta})
  = f_{\rm FD}(E_{\zeta}^{\eta}) \theta(E_{\zeta}^{\eta} + E_{c}) ,
\end{align}
just after the Matsubara summation is performed in Eq.~(\ref{eq:def-response}).
Accordingly, $\partial f_{\rm FD}(E_{\zeta}^{\eta})/\partial E_{\zeta}^{\eta}$
in Eq.~(\ref{eq:CME-stat-pre}) should be replaced with
\begin{align}
  \frac{\partial \tilde{f}_{\rm FD}(E_{\zeta}^{\eta})}
       {\partial E_{\zeta}^{\eta}}
  = -\delta(E_{\zeta}^{\eta})+\delta(E_{\zeta}^{\eta}+E_{c}) .
\end{align}
The second term induces a correction at the cutoff energy,
which is relevant in the system with an unbounded energy spectrum.
Consequently, Eq.~(\ref{eq:CME-stat-pre}) is modified to
\begin{align}
  \alpha^{\rm CME}_{\rm stat}
  & = e^{2}\int\frac{d^{3}k}{(2\pi)^{3}}\sum_{\zeta = R,L}\sum_{\eta=\pm}
      d_{\zeta}\mib{\Omega}_{\zeta}\cdot\nabla_{\mib k} E_{\zeta}^{\eta}
        \nonumber \\
  & \hspace{14mm}
      \times
      \bigl[ \delta(E_{\zeta}^{\eta})
            -\delta(E_{\zeta}^{\eta}+E_{c}) \bigr] ,
\end{align}
where the irrelevant term
with $\partial\Gamma_{\zeta}/\partial k_{z}$ is ignored.
The correction (i.e., the second term in the square brackets)
cannot be captured in an ordinary cutoff procedure.
The above expression is rewritten as
\begin{align}
  \alpha^{\rm CME}_{\rm stat}
  & = e^{2}\int\frac{d^{3}k}{(2\pi)^{3}}
      \sum_{\eta=\pm}
      \biggl[
      \frac{\eta v^{3}}{2d_{R0}}
      \left[ \delta(E_{R0}^{\eta})-\delta(E_{R0}^{\eta}+E_{c}) \right]
            \nonumber \\
  & \hspace{14mm}
    - \frac{\eta v^{3}}{2d_{L0}}
      \left[ \delta(E_{L0}^{\eta})-\delta(E_{L0}^{\eta}+E_{c}) \right]
      \biggr]
\end{align}
with $E_{R0}^{\eta} = b_{0}-\mu +\eta d_{R0}$
and $E_{L0}^{\eta} = -b_{0}-\mu +\eta d_{L0}$,
where
\begin{align}
 d_{R0} & = v\sqrt{k_{x}^{2}+k_{y}^{2}+(k_{z}-k_{0})^{2}} ,
            \\
 d_{L0} & = v\sqrt{k_{x}^{2}+k_{y}^{2}+(-k_{z}-k_{0})^{2}} .
\end{align}
We readily find that $\alpha^{\rm CME}_{\rm stat} = 0$.
This indicates that the Weyl model properly describes the CME
if an energy cutoff is appropriately taken into account.
Note that the energy cutoff gives the correction equivalent to
the anomaly contribution,
\begin{align}
  \alpha^{\rm CME}_{\rm stat, an}
  = \frac{e^{2}}{2\pi^{2}}b_{0} ,
\end{align}
arising from the ultraviolet limit.~\cite{takane1}
That is, the energy cutoff plays the same role as the regularization scheme
of Ref.~\citen{takane1} proposed on the basis of a quantum anomaly
in relativistic field theory.~\cite{fujikawa}

Under the energy cutoff, Eq.~(\ref{eq:CME-stat}) is justified
for the Weyl model as the surface term $c_{S}$ completely vanishes.
Note that the term equivalent to Eq.~(\ref{eq:CME-stat})
has been derived in a semiclassical theory.~\cite{son1,son2,basar}
Our argument microscopically justifies the semiclassical derivation
of Eq.~(\ref{eq:CME-stat}) based on the Weyl model.

\section{Summary}

We proposed a regularized continuum model that can describe
the anomalous electromagnetic response of a Weyl semimetal
[i.e., anomalous Hall effect (AHE) and chiral magnetic effect (CME)]
in a unified manner.
Considering the analysis based on this model,
we show that the original Weyl model can properly describe the CME
if an energy cutoff procedure is applied in a careful manner,
although it fails to describe the AHE
without including the mass terms with $\sigma_{z}\Lambda$.

To properly describe the anomalous electromagnetic response, a continuum model
with a pair of energy valleys should satisfy two requirements.
Concerning the AHE, the model must satisfy the requirement that
a winding number defined in each valley takes an integer and
the sum of the winding numbers for two valleys is
identified with the Chern number.
The regularized continuum model satisfies this requirement owing to
the presence of the mass terms, whereas the Weyl model does not satisfy it.
Concerning the CME, the requirement is that
the Berry curvature is conserved in reciprocal space,
in addition to the convergence of integration over $\mib{k}$.
Several parameters of the regularized continuum model are
adjusted to satisfy this requirement.
Contrastingly, the Weyl model automatically satisfies this,
although the unbounded spectrum should be regularized for convergence.
It is shown that the Weyl model can properly describe the CME
if an energy cutoff procedure is appropriately applied.

\section*{Acknowledgment}

This work was supported by JSPS KAKENHI Grant Number JP18K03460.


\begin{thebibliography}{99}
\bibitem{comment1} If Eq.~(\ref{eq:par-alpha}) is applied to a lattice model,
$S$ should be replaced with the boundary of the first Brillouin zone.
\bibitem{comment2} We cannot simply set $f_{\rm FD}(E_{\zeta}^{+}) = 0$
and $f_{\rm FD}(E_{\zeta}^{-}) = 1$ if $\mu < -\Delta_{M}$
or $\Delta_{M} < \mu$.
For example, in the case of $\mu < -\Delta_{M}$, we observe that
$E_{\zeta}^{-}$ at $k_{z} = \pm k_{M}$ is not necessarily smaller than $\mu$
near $(k_{x},k_{y}) = (0,0)$
reflecting the breaking of the conical structure in the $k_{z}$-direction.
\end{thebibliography}
\end{document}